\documentclass[11pt,twocolumn]{article}

\newcommand{\beq}{\begin{equation}}
\newcommand{\eeq}{\end{equation}}
\newcommand{\ba}{\begin{eqnarray}}
\newcommand{\ea}{\nonumber \end{eqnarray}}
\newcommand{\be}{\begin{eqnarray}}
\newcommand{\ee}{\nonumber \end{eqnarray}}

\newcommand{\bs}{\begin{slide}}
\newcommand{\es}{\end{slide}}
\newcommand{\bc}{\begin{center}}
\newcommand{\ec}{\end{center}}
\newcommand{\bi}{\begin{enumerate}}
\newcommand{\ei}{\end{enumerate}}

\def\fun#1#2{\lower3.6pt\vbox{\baselineskip0pt\lineskip.9pt
\ialign{$\mathsurround=0pt#1\hfil##\hfil$\crcr#2\crcr\sim\crcr}}}

\begin{document}
\twocolumn[
\begin{center}\bf
THE RIDDLE OF THE \boldmath $f_0(980)$ AND $a_0(980)$: ARE THEY THE
QUARK--ANTIQUARK STATES?

\bigskip
A.V. Anisovich, V.V. Anisovich, L.G. Dakhno, V.N. Markov,\\ M.A.
Matveev, V.A. Nikonov, A.V. Sarantsev  \\
{\it Petersburg Nuclear Physics Institute RAS, Russia}

\end{center}

\bigskip
]

The nature of mesons $f_0(980)$ and $a_0(980)$ is of principal meaning
for the systematics of scalar states and search for exotic mesons. This
is precisely why up to now the people lively discuss the problem of
whether the mesons $f_0(980)$ and $a_0(980)$ are the lightest scalar
quark--antiquark particles or they are exotic states,
like  four-quark ($q\bar q q\bar q$) states \cite{Jaffe},
$K\bar K$ molecule \cite{Isgur} or minions \cite{Close} -- see,
for example review \cite{Bugg} and references therein.

Investigations of the PNPI group [4--13] favoured the opinion that
$f_0(980)$ and $a_0(980)$ are dominantly $q\bar q$ states, with a small
(10--20\%) admixture of the $K\bar K$ loosely bound component. There
exist both qualitative arguments and calculations of certain reactions
that support this idea. First, let us discuss the qualitative
arguments.

$(i)$ In hadronic reactions, the resonances $f_0(980)$ and
$a_0(980)$ are produced as standard, non-exotic resonances,
with compatible yields and similar distributions. This was observed in
the meson central production at the high energy  hadron--hadron
 collisions (data of GAMS \cite{GAMS} and Omega
 \cite{Omega} Collaborations) or in hadronic decays
of $Z^0$ mesons (OPAL Collaboration \cite{OPAL}).

$(ii)$ The supposed exotics of   $f_0(980)$ and $a_0(980)$ was often
argued relying on a surprising proximity of their masses, while it
would be natural to expect the variation of masses in the nonet to be
100--200\,MeV. Note that Breit--Wigner resonance pole,
which determines a true mass of the state, is rather sensitive to a
small admixture of hadron components, if the production threshold for
these hadrons lays nearby. As to  $f_0(980)$ and $a_0(980)$, it is easy
to see that a small admixture of the $K\bar K$ component shifts the
pole to the $K\bar K$ threshold independently of whether it is above or
below threshold. Besides, the peak observed in the main mode of the
 $f_0(980)$ and $a_0(980)$ decays,  $f_0(980)\to \pi\pi$ and
$a_0(980)\to \eta\pi$, is always slightly  below the $K\bar K$
threshold that mimics Breit--Wigner resonance with the mass below 1000
MeV ($K\bar K$ threshold). This imitation of a resonance created the
legend about a ``surprising proximity" of the $f_0(980)$ and $a_0(980)$
masses.

In fact, the mesons $f_0(980)$ and $a_0(980)$ are characterised by not
one pole (as is the Breit--Wigner case) but two poles (as in
Flatt\'e formula or $K$-matrix approach), and these poles are rather
different for $f_0(980)$ and $a_0(980)$ [4,5]. Note that
Flatt\'e formula is unable to give us adequate description of spectra
near these poles. So we should apply more complicated
representation of the amplitude \cite{1} or the $K$ matrix approach
\cite{2}, see also \cite{Au}.

{\bf In parallel with the above mentioned qualitative
considerations, there exist convincing arguments which favour the
quark--antiquark nature of $f_0(980)$ and $a_0(980)$:}\\
{\large\bf
(I)} Systematics of mesons on linear trajectories on the $(n,M^2)$ and
$(J,M^2)$ planes ($n$ is the radial quantum number of meson $M$
and $J$ its spin)   support quark--antiquark structure of $f_0(980)$ and
$a_0(980)$ [6]. The states which can be put
 on  linear $(n,M^2)$- and $(J,M^2$)-trajectories
are quark--antiquark states. {\it Vice versa}:
the states which do not belong to linear trajectories should be
considered as exotics.
It occurred that
   $f_0(980)$ and $a_0(980)$ lay
comfortably on linear $q\bar q$ trajectories and, together with
$f_0(1300)$
and $K_0(1415)$, they form the lowest
scalar nonet with rather good accuracy. Also, on the basis
of this systematics, the second ($n=2$) scalar nonet was reconstructed:
$f_0(1500)$, $f_0(1750)$, $a_0(1520)$, $K_0(1850)$ [5,7-9].\\
{\large\bf (II)}
The $K$-matrix analysis has been carried out in [5,9] as well as
in a series of precedent papers for the states with isotopic spins
$I=0,1,1/2$, i.e., for $f_0$, $a_0$ and $K_0$ mesons. Such analysis
allowed us to reconstruct the characteristics of real mesons and
so-called "bare states" -- the states which precede real mesons before
switching on the decay channels.
The performed analysis allowed us
to construct two first scalar nonets also in terms of bare states and to
trace the tranformation of corresponding poles by switching on
gradually the decay processes. The resonances, which are the
descendants of bare $q\bar q$ mesons with
 $n=1$, are
$a_0(980)$, $f_0(980)$, $f_0(1300)$, $K_0(1430)$ and those with $n=2$
are
$a_0(1520)$, $f_0(1500)$, $f_0(1750)$, $K_0(1850)$.
The $K$-matrix analysis points to the
existence of a broad resonance $f_0(1200-1600)$, which does not belong
to linear trajectory, thus being the candidate for exotic state.
The decay constants of this broad states, $f_0(1200-1600)\to
\pi\pi,K\bar K,\eta\eta,\eta\eta'$, are compatible with the glueball
nature of this resonance.
 According to this
analysis, $a_0(980)$ and $f_0(980)$ are dominantly the quark--antiquark
states and  $f_0(980)$ has a considerable $s\bar s$ component, of the
order of 60--70\%. The results of this analysis are summarised in the
review [8] and Chapter V of the monograph \cite{7}.\\
{\large\bf (III)}
Hadronic decay of the $D^+_s$-meson, $D^+_s\to \pi^+f_0(980)
\to \pi^+\pi^+\pi^-$:
on the quark level, the decay  goes as $c\bar s \to \pi^+ s\bar s\to
 \pi^+f_0(980)$ just proving the dominance of the $s\bar s$
 component in $f_0(980)$. Our analysis \cite{8}
showed that
2/3 $s\bar s$ is contained in $f_0(980)$, and this estimate is
supported by the experimental value:
 $BR\left (\pi^+f_0(980)\right)= 57\%\pm 9\%,$
and 1/3 $s\bar s$ is dispersed over the resonances
$f_0(1300)$, $f_0(1500)$, $f_0(1200-1600)$.
    So the reaction $D^+_s \to \pi^+ +f_0$ is
a measure of the $1^3P_0s\bar s$ component in the $f_0$ mesons, it
definitely tells us about the dominance of the $s\bar s$ component in
$f_0(980)$, in accordance with the results of the $K$-matrix
analysis. The conclusion ab
out dominance of $s\bar s$ component in
$f_0(980)$  was also made on the basis of the analysis
of the decay $D^+_s\to \pi^+\pi^+\pi^-$ in papers
\cite{Gatto,Rupp,Ochs}.\\
{\large\bf (IV)} Radiative decays
  $f_0(980)\to \gamma\gamma$, $a_0(980)\to \gamma\gamma$
 agree well with the
 calculations \cite{9} based on the assumption of the
 quark--antiquark
nature
of these mesons. Let us emphasize again that the calculations favour
the $s\bar s$ dominance in $f_0(980)$. \\
{\large\bf (V)} Radiative decay $\phi(1020)\to \gamma f_0(980)$
was a subject of lively discussion in the latest years: there existed
an opinion that this decay strongly contradicts the hypothesis of its
$q\bar q$ nature \cite{NNA,FEC,DW}. However, our calculations
\cite{9,10} carried out within both relativistic and nonrelativistic
approaches showed that the $q\bar q$ nature of $f_0(980)$
agrees well with data. In \cite{10}, we focus precisely on the
problems, that  appear when Siegert's theorem \cite{Sieg}  is
straightforwardly applied
 to the decay $\phi(1020)\to \gamma f_0(980)$,
though  the produced $f_0(980)$ is characterized by two poles in
the complex-mass plane.

Invesigations, which are discussed in items (II)--(IV), are based on the
use of the covariant moment-expansion technique developed in 90's by
the PNPI group in a number of papers and summarized in \cite{operator}.

Correct systematics of the lowest scalar $q\bar q$ states is a basis
for a reliable search for the glueballs.  Precisely this systematics
allowed us to identify the broad $f_0(1200-1600)$ state as the lowest
scalar glueball. Now we have strong indication to the existence of
tensor glueball \cite{11}, thus opening a new page of physics ---
the glueball physics.

\end{document}